\documentclass{aa}
\usepackage[english]{babel}
\usepackage{latexsym}
\usepackage{epsfig}
\usepackage{psfig}
\usepackage{amssymb}
\usepackage{amsfonts}
\usepackage{graphicx}
\usepackage{natbib}

\defcitealias{BL99}{BL}
\bibpunct{(}{)}{;}{a}{,}{,}

\begin{document}

\title{Non-Keplerian rotation in the nucleus of \object{NGC 1068}:
evidence for a massive accretion disk?}

\titlerunning{Massive accretion disk in NGC 1068}

\author{G. Lodato \inst{1} \and G. Bertin \inst{2}}

\institute{Scuola Normale Superiore, Piazza dei Cavalieri
7, I-56126, Pisa, Italy \and Universit\`a degli Studi di Milano,
Dipartimento di Fisica, via Celoria 16, I-20133 Milano, Italy}

\date{Received date/Accepted date} 

\abstract{The nucleus of the Seyfert galaxy NGC 1068 is believed to
host a supermassive black hole.  Evidence for the presence of a
massive central object is provided by water maser emission, which
displays a linear pattern in the sky, suggestive of a rotating
disk. The rotating disk hypothesis is further strengthened by the
declining shape of the derived rotation curve. Similar maser emission
from NGC 4258 has led to a reliable estimate of the mass of the
central black hole, because in this case the rotation curve is
Keplerian. In the case of NGC 1068 the rotation curve traced by the
water maser is non-Keplerian. In this paper we provide an
interpretation of the non-Keplerian rotation in NGC 1068 by means of a
self-gravitating accretion disk model. We obtain a good fit to the
available data and derive a black hole mass $M_{\bullet}\approx
(8.0\pm 0.3)~10^6M_{\sun}$. The resulting disk mass is comparable to
the black hole mass. As an interesting by-product of our fitting
procedure, we are able to estimate the viscosity parameter, which
turns out to be $\alpha\approx 10^{-2}$, in line with some theoretical
expectations.  }

\maketitle

\section{Introduction}

There is now convincing evidence that most AGNs host a supermassive
black hole, with masses ranging from $10^7M_{\sun}$ to
$10^9M_{\sun}$. The gravitational energy extracted from an accretion
disk around such black holes is generally considered to be the main
source of the AGN luminosity. Determining the mass of the central
black hole $M_{\bullet}$ is thus one important goal of studies of
AGNs. This has received greater attention recently, in a more general
context, because correlations have been found between $M_{\bullet}$
and the global properties of the host galaxy, such as its mass or
luminosity \citep{magorrian98} or its central velocity dispersion
\citep{ferrarese2000,gebhardt2000}.

The central point mass $M_{\bullet}$ can be measured by different
methods. The analysis of {\it HST} optical spectra in terms of stellar
dynamical models \citep[see, for example,][]{gebhardt2000b} and the
study of gas kinematics from {\it HST} spectra \citep{vandermarel98}
probe the nuclear gravitational field of nearby galaxies at distances
typically of the order of $\approx 100$ pc from the
center. $M_{\bullet}$ can be estimated also from reverberation mapping
of the broad line region, which is able to probe the gravitational
field at smaller scales. However, reverberation mapping is not
applicable to Type II Seyfert galaxies, like NGC 1068, in which the
broad line region is hidden from our line of sight. An independent
powerful tool to obtain reliable estimates of $M_{\bullet}$ is
provided by the study of the Doppler shift of water maser emission
lines (when available). This latter method provides the most reliable
determinations of $M_{\bullet}$, because it probes the gravitational
field at very small distances (less than 1 pc) from the center. The
masing spots are often observed to trace a linear structure and show a
declining rotation curve. In the case of \object{NGC 4258}
\citep{miyoshi95}, the rotation curve is remarkably Keplerian and
leads to a robust determination of $M_{\bullet}\approx
3.9~10^7M_{\sun}$.

\object{NGC 1068} is one of the best studied Seyfert galaxies. It is
considered to be the prototypical Type II Seyfert, in which the
central engine is believed to be hidden from our line of sight by a
dusty structure. This obscuring structure, now resolved with VLBA
radio continuum observations \citep{gallimore97}, appears to have a
disk-like rather than a toroidal shape. The bolometric luminosity of
the nucleus is $L_\mathrm{bol}\approx 8~10^{44}$ erg/sec
\citep{pieretal94}. If we adopt for the accretion efficiency a value
of $\eta=0.06$, appropriate for a non-rotating black hole, the
resulting mass accretion rate is $\dot{M}\approx 0.235~
M_{\sun}/\mbox{yr}$. Throughout this paper we will assume that the
distance to NGC 1068 is 14.4 Mpc (so that $1\mbox{ mas}=0.069\mbox{ pc}$).

Water maser emission is observed from the nucleus of NGC 1068
extending out to a radius $r\approx 1$ pc
\citep{greenhill96,greenhill97}. This emission is believed to come
from a rotating, almost edge-on accretion disk, possibly associated
with the disk seen in VLBA continuum \citep{gallimore97}. The striking
feature of this maser emission is that, in contrast with the case of
NGC 4258, the rotation curve of the masing spots is clearly {\it
non-Keplerian}. \citet{greenhill96} report a best fit to the data with
a circular velocity $V\propto r^{-0.31}$. Possible causes for the
sub-Keplerian rotation may be: a) the maser may not be associated with
a pure rotating structure, because an outflow is present; b) radiation
pressure may reduce the rotational velocity \citep{pierkrolik92a}; c)
the source of the gravitational field may be extended, including a
nuclear stellar cluster (as proposed by \citealt{kumar99}) or the
accretion disk itself.

In this paper we interpret the non-Keplerian rotation in NGC 1068 in
terms of a self-gravitating accretion disk model (\citealt[ hereafter
BL]{BL99}). The disk is assumed to be self-regulated at the threshold
of Jeans instability. The gravitational field is computed by solving
the relevant Poisson equation including both the central point-like
object and the disk. The resulting rotation curve is characterized by
an inner Keplerian curve connected with an outer asymptotically flat
rotation curve.

We have fitted the Very Long Baseline Interferometry (VLBI) data of
\citet{greenhill97} with our theoretical models. The fit is very
good. From the results of the fit we can derive the value of the
central black hole mass, $M_{\bullet} \approx 8~10^6 M_{\sun}$,
roughly one half of the value that would be inferred from a Keplerian
fit (from which one would obtain $M_{\bullet}\approx 1.5~10^7
M_{\sun}$), basically because we attribute part of the gravitating
mass to the disk. As an interesting by-product of our modeling
procedure, we are able to derive the value of the viscosity parameter
$\alpha$ that regulates the accretion process, which turns out to be
in agreement with some theoretical expectations.

The paper is organized as follows. In Section \ref{preliminary} we
describe the role played by the disk self-gravity in the outer parts
of AGNs. In Section \ref{ngc} we report the observations of the
nucleus of NGC 1068, focusing on its geometry and kinematics. In
Section \ref{fit} we describe our fit to the water maser data. In
Section \ref{alt} we discuss some possible alternatives to the model
presented in this paper. In Section \ref{conclusion} we draw our
conclusions. In Appendix \ref{likelihood} we describe in some detail
the statistical significance of the fitting procedure adopted.

\section{Preliminary considerations on the influence of the disk
self-gravity in AGN accretion disks}
\label{preliminary}

The disk self-gravity may affect several aspects of the dynamics of
accretion disks: a) gravitational instabilities are expected to modify
the energy and angular momentum transport in the disk, perhaps being
the main tool able to drive accretion at large radii \citep{lin87}; b)
the vertical gravitational field associated with the disk modifies the
vertical hydrostatic equilibrium \citep{pacinski78,bardou}; c) the
radial gravitational field of the disk may lead to deviations from
Keplerian rotation \citepalias{BL99}.

\subsection{Self-regulation of Jeans instability}

The onset of gravitational instabilities in a fluid disk is
controlled by the well-known axisymmetric stability parameter $Q$:
\begin{equation}
\label{Q}
Q=\frac{c_s\kappa}{\pi G\sigma},
\end{equation}
where $c_\mathrm{s}$ is the thermal speed, $\kappa$ is the epicyclic frequency,
and $\sigma$ is the disk surface density. For $Q<1$ the disk is
unstable. As $Q$ is proportional to the thermal speed, a hot enough
disk is expected to be free from the effects related to the disk
self-gravity, while, on the other hand, if the disk is cold enough to
begin with, it cannot survive long in such a condition and is likely
to be eventually characterized by a value of $Q$ close to unity, as a
result of a self-regulation mechanism, studied and recognized
especially in the field of galactic dynamics
\citep[see][]{bertinbook}.  Self-regulation has been sometimes taken
into account also in the context of accretion disks
\citep{lin87,hure2000}. The inner parts of the accretion disk in AGNs
are very hot, so that the disk self-gravity is not expected to play a
role there. On the other hand, it is easy to show that the outer
colder parts of the disk may be subject to gravitational
instabilities. Consider the outer region of the $\alpha$-disk solution
by \citet{shakura73}, for the case in which gas pressure and free-free
absorption dominate. The radial dependence of $Q$ in this case is:
\begin{equation}
\label{qprof}
Q\approx 5.6~10^3\alpha^{7/10}\dot{M}_{26}^{-11/20}M_8^{-3/4}\hat{r}^{-9/8},
\end{equation}
where we have scaled the basic physical parameters to typical AGN
values, so that $\dot{M}_{26}$ is the mass accretion rate in units of
$10^{26}$ g/sec $\approx 1.57 M_{\sun}$/yr, $M_8$ is the black hole
mass in units of $10^8M_{\sun}$, and $\hat{r}$ is the radius in units
of the Schwarzschild radius of the black hole
$R_{\bullet}=2GM_{\bullet}/c^2$; $\alpha$ is the dimensionless
viscosity parameter entering the Shakura-Sunyaev prescription for
viscosity. The value of $Q$ thus decreases rapidly with increasing
radius and becomes equal to unity at $\hat{r}_Q=r_Q/R_{\bullet}\approx
10^3$ (for $M_8=\dot{M}_{26}=1$ and $\alpha=0.01$), i.e. at
$r_Q\approx 10^{-2}$ pc. \citet{kumar99} has carried out a similar
analysis in the case in which the opacity is dominated by metal grains
and has basically confirmed the above estimate, finding that $Q\approx
1$ at $r_Q\approx 3~10^{-3}$ pc, for the same input parameters.

The argument of self-regulation suggests that the outer disk, beyond
$r_Q$, be characterized by $Q\approx 1$.

\subsection{Impact of the disk self-gravity on the vertical structure}

If we consider the modifications of the vertical hydrostatic
equilibrium by the disk self-gravity, a simple way to address the
problem is to compare the vertical lengthscales derived in the
limiting cases of non-self-gravitating disk and of fully
self-gravitating disk. In the non-self-gravitating case we have:
\begin{equation}
h_{\mathrm{nsg}}=\frac{c_s}{\Omega},
\end{equation}
while in the self-gravitating case:
\begin{equation}
h_{\mathrm{sg}}=\frac{c_s^2}{\pi G\sigma}.
\end{equation}
The disk self-gravity will then modify the vertical structure of the disk
if:
\begin{equation}
\frac{h_{\mathrm{sg}}}{h_{\mathrm{nsg}}}=\frac{c_s\Omega}{\pi G\sigma}
\approx Q\approx 1,
\end{equation}
because $\kappa$ is related to $\Omega$ by a numerical factor
(dependent on the rotation curve) close to unity. Therefore, a
discussion of the vertical equilibrium also leads to the conclusion
that an accretion disk is expected to be self-gravitating if it
extends to radii larger than $r_Q$. This simplified analysis is
confirmed by the more detailed study presented earlier by us
\citepalias[ Appendix]{BL99}, where a convenient analytical expression
for the disk thickness, in the case in which both the disk and the
central object are taken into account, is provided:
\begin{eqnarray}
\nonumber h= & \displaystyle \frac{c_s^2}{\pi
G\sigma}\frac{\pi}{4Q^2(2\Omega^2/\kappa^2-1)}\times\\
   &
\left[\sqrt{\displaystyle 1+\frac{8}{\pi}Q^2\left(\frac{2\Omega^2}{\kappa^2}
-1\right)}-1\right].
\end{eqnarray}

\subsection{Modification of the rotation curve}

In this paper we will refer to the self-gravitating steady-state disk
model of \citetalias{BL99}, who considered disks that are
self-regulated with respect to Jeans instability, so that
$Q=\bar{Q}\approx 1$. The gravitational field of the disk is computed
by solving self-consistently the relevant Poisson equation (see
Eq. (4) of \citetalias{BL99}). 

The basic dynamical feature of such self-regulated disk model is that
at large radii the rotation curve $V(r)$ departs from the Keplerian
profile, approaching an asymptotically flat rotation curve. The
typical lengthscale that marks the transition from Keplerian to
self-gravitating regime is:
\begin{equation}
\label{rs}
r_{\mathrm s}=2GM_{\bullet}\left(\frac{\bar{Q}}{4}\right)^2
\left(\frac{G\dot{M}}{2\alpha}\right)^{-2/3}.
\end{equation}
Deviations from Keplerian rotation occur already at radii of the order
of a fraction of $r_{\mathrm s}$; in fact, we can have $\mbox{d ln}
V/\mbox{d ln} r=-0.4$ at $r=0.1r_{\mathrm s}$. For $r\gg r_{\mathrm
s}$, the mass of the disk grows linearly, and the surface density
behaves as $\sigma\propto 1/r$. If we estimate the value of
$r_{\mathrm s}$, assuming the typical AGN values for the relevant
parameters used earlier ($M_8=\dot{M}_{26}=1$ and $\alpha=0.01$), we
find that $r_{\mathrm s}$ is of the order of a few pc. Interestingly,
we find that deviations from Keplerian rotation are expected to occur
exactly at a distance from the center probed, in the case of NGC 1068,
by the water maser emission. In contrast, in the case of NGC 4258, a
similar analysis would lead to an estimated $r_{\rm s}\gg 1$ pc,
consistent with the Keplerian rotation curve observed by
\citet{miyoshi95}.

\begin{figure*}
  \includegraphics[width=12cm]{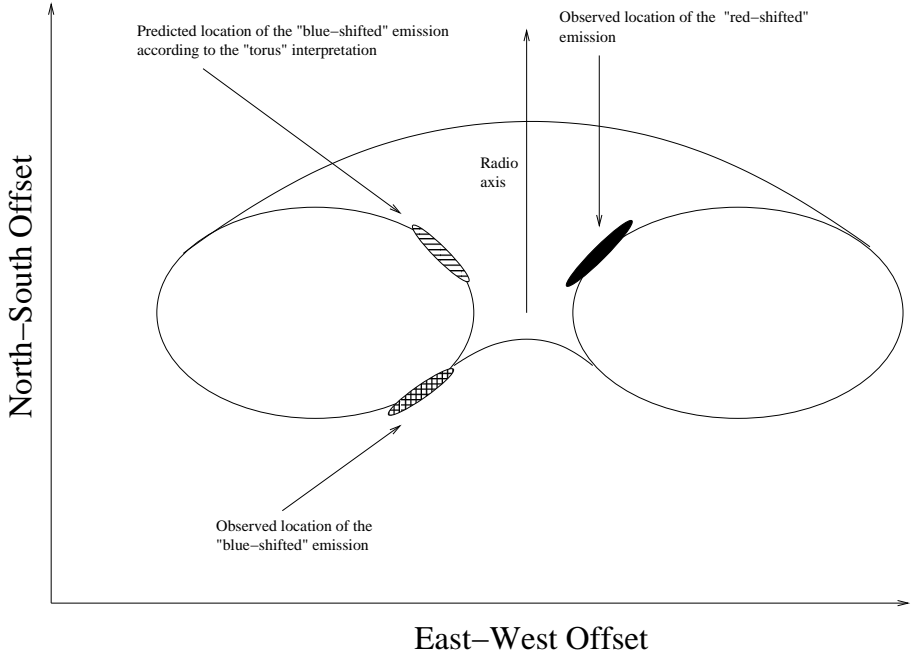}
  \caption{Schematic representation of the location of the maser
  emission according to the torus interpretation. The three elliptical
  patches show the location of the observed ``red-shifted" masers, of the
  predicted ``blue-shifted" emission, based on the torus interpretation,
  and of the observed ``blue-shifted" emission (see also
  \citealt{greenhill96}).}  \label{fig:torus}
\end{figure*}

Based on the arguments presented in this Section, we therefore
conclude that, if the water maser emission from NGC 1068 traces the
outer parts of the nuclear accretion disk, such disk is likely to be
self-gravitating for what concerns its vertical structure and
transport phenomena, and it should display significant deviations
from Keplerian rotation.

\section{The masing spots in \object{NGC 1068}: geometry and
kinematics}
\label{ngc}

In this Section we will discuss the geometry of the maser emission in
NGC 1068, for which different interpretations have been given. 

\subsection{Disk or torus?}

\citet{greenhill96} observed the ``red-shifted" water maser emission
from NGC 1068, finding that it traces a linear pattern in the plane of
the sky, inclined by approximately $45\degr$ with respect to the
direction of the radio jet. The misalignment between the radio axis
and the maser emission led to an early interpretation of the maser
spots as arising from the upper limb of a nearly edge-on torus,
characterized by a rather large aspect ratio (see the schematic
representation in Fig. \ref{fig:torus}). According to this
interpretation, we would expect that the corresponding ``blue-shifted"
emission should come from the dashed region shown in Fig.
\ref{fig:torus}. On the other hand, subsequent observations
\citep{greenhill97} showed that the emission traces a linear pattern
from the ``red-shifted" to the ``blue-shifted" emission, hence arguing
in favor of a thin disk interpretation. The misalignment between the
disk axis and the radio jet is not uncommon in AGNs
\citep[see][]{schmitt02} and may be due to a variety of physical
mechanisms (for example, a warp in the outer disk;
\citealt{pringle97}).

\subsection{Geometry of the disk emission}
\label{geometry}
The rotation curve of the maser spots can be divided in two different
regions: 1) at small impact parameter the masers show an apparently
rising rotation curve; 2) starting from a radial distance $\approx
0.6$ pc from the center the rotation curve declines, following a
sub-Keplerian profile.

The natural interpretation of the declining part of the rotation curve
is that it arises from material that moves parallel to our line of
sight (i.e. that lies on a disk diameter perpendicular to the line of
sight).  The best argument in favor of this interpretation is that
maser amplification is largest for material that lies close to the
line of the nodes. On the other hand, the rising part of the observed
``rotation curve'' is thought to originate from one quarter of the
disk at the inner maser disk radius, so that the rising curve is an
effect of velocity projection along the line of sight \citep[see
also][]{miyoshi95}. According to this interpretation, the inner
radius of the maser disk is located at $\approx 0.6$ pc and the outer
disk radius is at $\approx 1$ pc.

\citet{baan96} claimed to have observed a drift in the velocity of the
water masers, indicating that the maser spots are subject to large
accelerations, incompatible with the disk interpretation, according to
which the maser centripetal acceleration should be perpendicular to
the line of sight. Those large accelerations have not been confirmed
by subsequent work by \citet{gallimore01}, who, monitoring the
velocity drift, find that the maser spots between 0.6 pc and 1 pc
should lie within $\theta\lesssim 2\degr$ from the line of nodes.

In the following we will therefore assume that the water maser
emission comes from an edge-on thin disk extending from 0.6 pc to 1
pc and that the declining part of the rotation curve comes from
material that lies within $2\degr$ from the line of nodes.

\section{Non-Keplerian rotation as an effect of the disk self-gravity}
\label{fit}

\subsection{Fit procedure and results}
\label{procedure}

Here we will fit the kinematical data of NGC 1068 by using the
completely self-regulated disk model of \citetalias{BL99} (in which,
for simplicity, the self-regulation prescription is taken to hold at
all radii). Actually, we expect the inner disk to be hotter and
characterized by a higher value of the stability parameter $Q$, so
that a partially self-regulated disk model (also described in
\citetalias{BL99}) would be more appropriate. In Subsection
\ref{consistency} we will justify the consistency of the simpler model
adopted below.

\begin{figure*}
  \includegraphics[width=12cm]{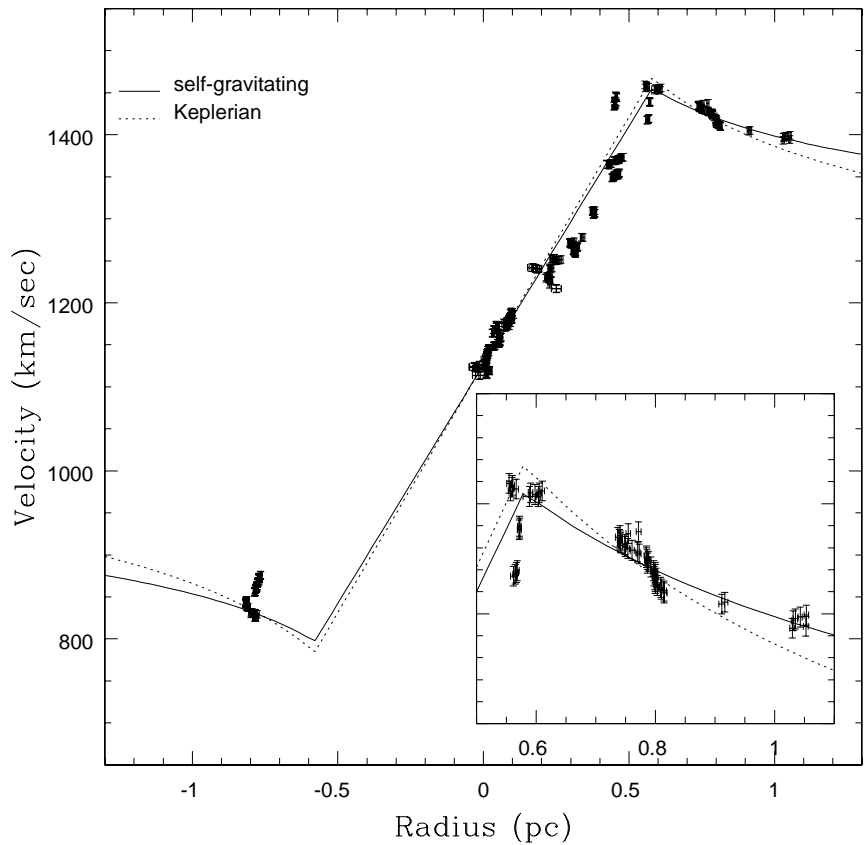}
  \caption{Fit to the rotation curve from the water maser emission by
  a self-gravitating accretion disk model. Data from
  \citet{greenhill97}. The small panel shows a blow up of the
  declining part of the rotation curve together with the best fit
  obtained by assuming Keplerian rotation. The error bars reported
  here include uncertainties beyond the actual VLBI spectral resolution
  (see text).}  
  \label{fig:fit}
\end{figure*}

The rotation curve of the completely self-regulated model by
\citetalias{BL99} is determined when one specifies the radial
lengthscale $r_{\mathrm s}$ (defined in Eq. (\ref{rs})), the velocity
scale $V_0$, defined as:
\begin{equation}
\label{V0}
V_0^2=\frac{8}{\bar{Q}^2}\left(\frac{G\dot{M}}{2\alpha}\right)^{2/3},
\end{equation}
and a dimensionless parameter (called $\xi$), which is proportional to
the net angular momentum flux in the disk $\dot{J}$. We will fix this
parameter by requiring the no-torque condition at the radius
corresponding to the innermost stable orbit around the black
hole. However, the specific choice of $\xi$ is not critical for our
conclusions, because we are exploring the disk properties at large
radii, where the effects of the inner boundary condition are
negligible. Therefore, we are left with two free dimensional scales,
that are obtained by fitting the available data (from
\citealt{greenhill97}).

Note that, from the definitions of $r_{\mathrm s}$ and $V_0$, we have
that:
\begin{equation}
V_0^2=\frac{GM_{\bullet}}{r_{\mathrm s}}.
\end{equation}
Therefore, from the value of $r_{\mathrm s}$ and $V_0$ we can estimate
$M_{\bullet}$ and $\dot{M}/\alpha$. In addition, once these quantities
are specified, from the models of \citetalias{BL99} we can obtain the
disk mass.

In our fit, we have restricted our attention to the data of the
declining part of the rotation curve of the ``red-shifted" maser
(i.e. the data points with $r> 0.6$ pc), because these data are
directly related to the gravitational potential. In contrast, the
velocity data points at $r< 0.6$, according to the interpretation of
the masing disk geometry presented in Subsection \ref{geometry}, only
reflect the rotation at the inner radius of the masing disk and do not
carry any additional information about the mass distribution. We have
assumed that the systemic velocity is $1126$ km/sec
\citep{greenhill97}. The uncertainty in the position of the maser
spots in the \citet{greenhill97} data is of the order of $\approx
50\mu$as. The spectral resolution of the VLBI data is $\lesssim 1$
km/sec. However, the velocity uncertainty $\Delta V$ is expected to be
higher due to the uncertainties in the estimate of the systemic
velocity and of the magnitude of turbulent motion in the disk. We will
assume that these uncertainties sum up to $\approx 10$ km/sec (see
also discussion in Appendix \ref{likelihood}). All fits are obtained
with a $\chi$-square minimization.  The result of the fit, shown in
Fig. \ref{fig:fit}, is satisfactory. The resulting reduced
$\chi$-square is $\tilde{\chi}^2=0.55$, with 48 degrees of
freedom. The fit parameters are $V_0=(110.4\pm 0.3)$ km/sec and
$r_{\mathrm s}= (2.82\pm 0.1)$ pc, where the uncertainties define the
68\% confidence level and are derived from the Hessian of the
$\chi$-square. The resulting black hole mass is $M_{\bullet}= (8.0\pm
0.3)~10^6M_{\sun}$ and the disk mass is approximately equal to the
black hole mass. We also obtain $\dot{M}=(28.1\pm 0.2)~\alpha
M_{\sun}/$yr. If we estimate $\dot{M}$ from the bolometric luminosity
(assuming an accretion efficiency $\eta=0.06$), we obtain
$\alpha\approx 8.3~10^{-3}$, a reasonable number.

The best-fit curve resulting from the self-gravitating disk model is
not a power-law. However, \citet{greenhill96} were able to obtain a
good fit to the data by assuming a rotation curve of the form
$V\propto r^{-0.31}$. If we compute the quantity $\mbox{d ln}
V/\mbox{d ln} r$ for our best fit model, we find that it ranges from
$-0.35$ at the inner edge of the disk to $-0.30$ at the outer edge. 

We have also performed a fit by assuming a Keplerian rotation
curve. The quality of the fit is definitely worse. The minimum
$\chi$-square in this case is in fact $\tilde{\chi}^2=1.557$ and is
formally rejected at the 95\% confidence level. The small panel of
Fig. \ref{fig:fit} compares the two models: the Keplerian best-fit
curve fails to reproduce both the highest and the lowest part of the
rotation curve. The resulting best-fit value of the black hole mass in
the Keplerian fit is $M_{\bullet}\approx (1.50\pm0.02)
~10^7M_{\sun}$. Note that the total mass (disk + black hole) of our
self-gravitating best fit model is roughly the same as the black hole
mass of the Keplerian fit. Therefore, a non-self-gravitating fit,
which attributes all the mass to the central object, gives the correct
value for the total mass, but fails to provide the correct slope of
the rotation curve.

One consequence of the smaller value of the black hole mass derived
from the self-gravitating disk model is that the corresponding
Eddington luminosity of the black hole is proportionally reduced,
hence leading to a higher Eddington ratio for the observed central
object. In fact, based on the results of our fit, we derive that
$L_\mathrm{bol}/ L_\mathrm{Edd}=0.77$, to be compared to the value
$L_\mathrm{bol}/ L_\mathrm{Edd}=0.41$ obtained from the Keplerian
fit. Note, however, that the Eddington luminosity is not well defined
when the mass is not spherically distributed.

\begin{figure*}
  \centerline{ \epsfig{figure=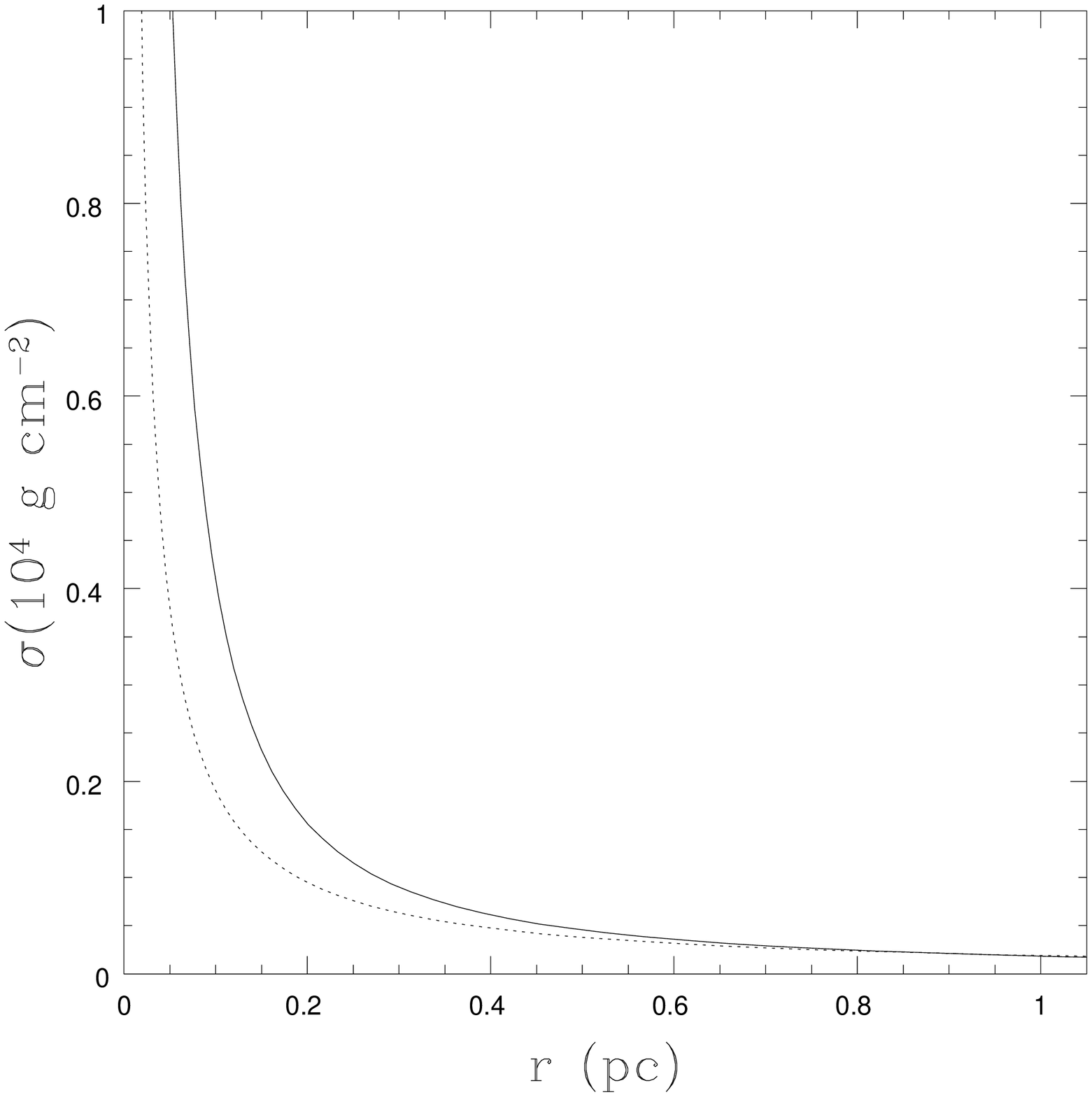,width=7.5cm}
  \epsfig{figure=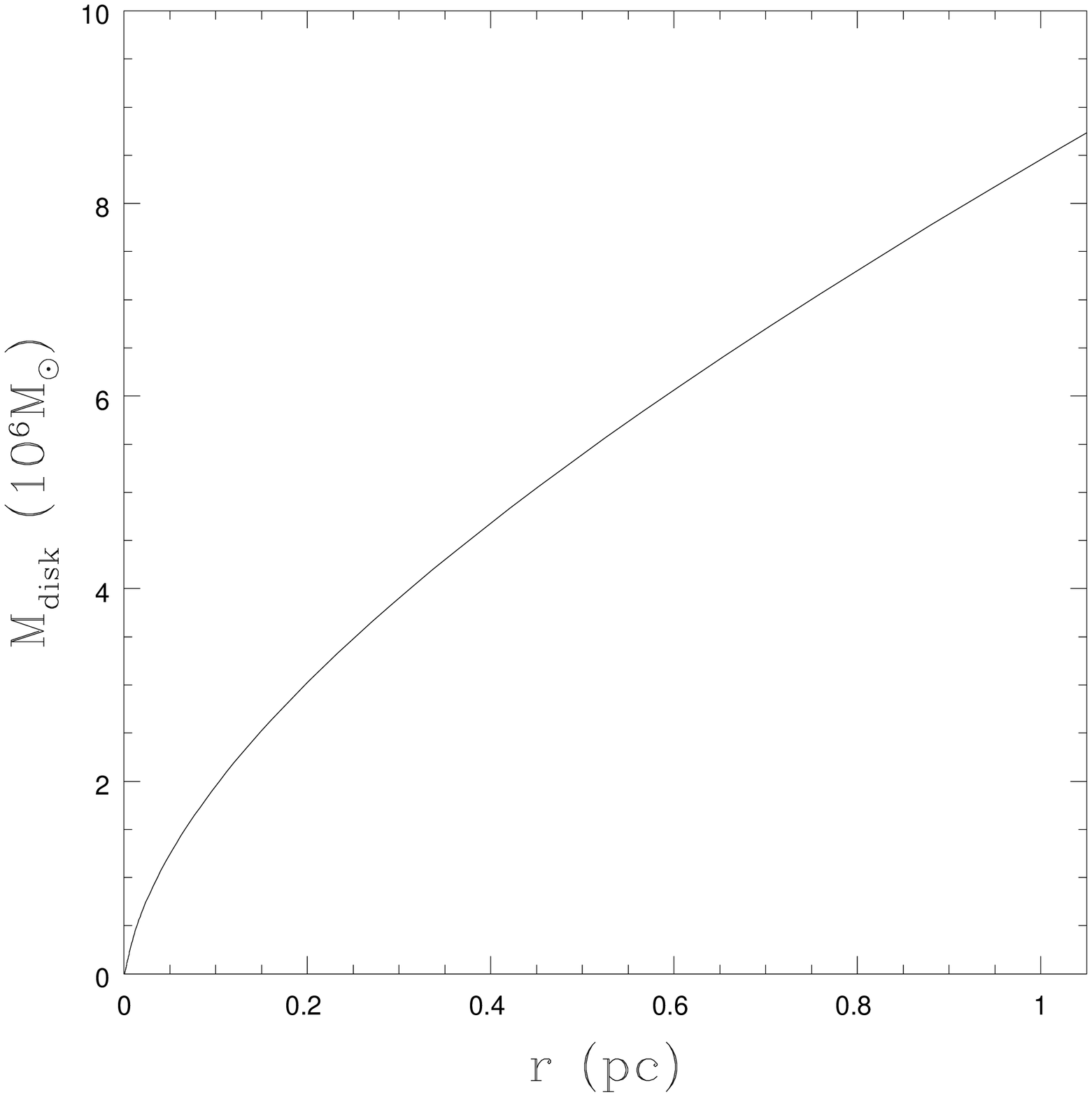,width=7.5cm}} \centerline{
  \epsfig{figure=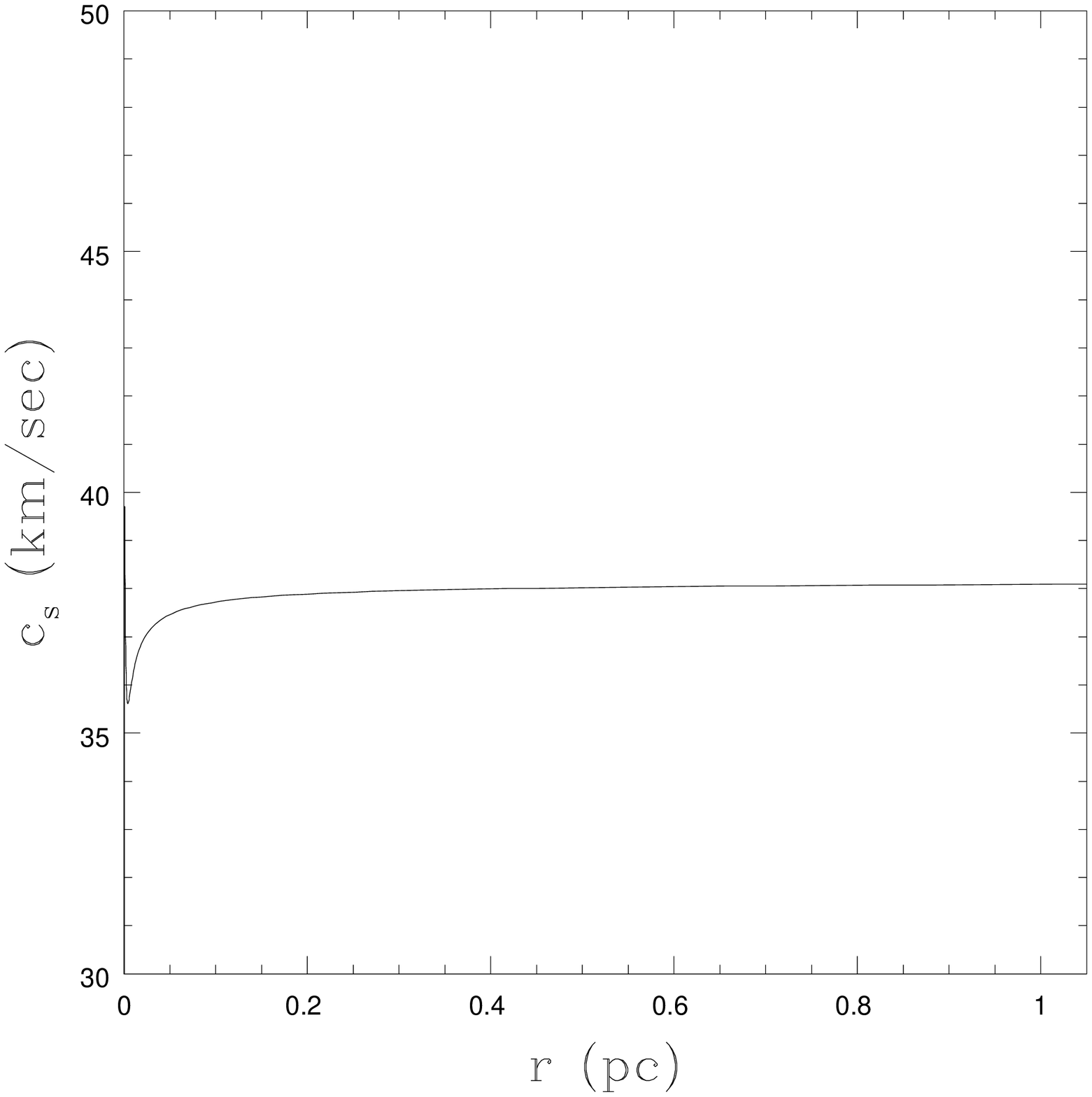,width=7.5cm}
  \epsfig{figure=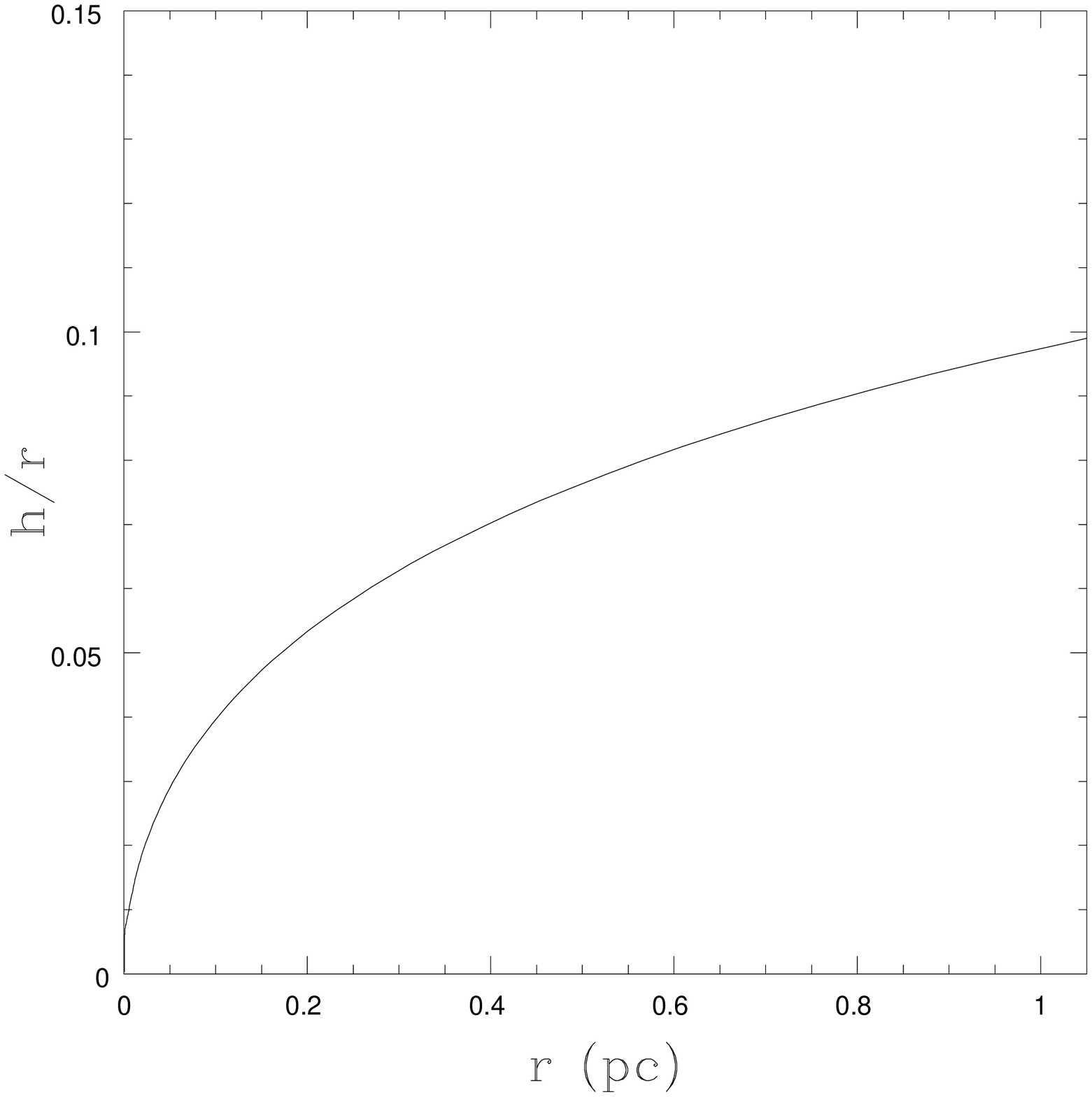,width=7.5cm}} 
  \caption{Surface density profile
  ({\it upper left}), cumulative mass ({\it upper right}), equivalent
  thermal speed ({\it lower left}), and aspect ratio ({\it lower
  right}) of the best-fit accretion disk model to the maser data of
  NGC 1068. The dotted line in the surface density plot is the curve
  with $\sigma\propto 1/r$ that matches the asymptotic behavior at
  large radii.}  
  \label{fig:properties}
\end{figure*}

Some data points (at approximately 0.8 pc from the center) are not
well fitted by neither the self-gravitating nor the Keplerian
model. The velocities of these points display a higher slope with
respect to that predicted by the best fit model. This discrepancy
could be due to the fact that the corresponding masing spots lie on a
spiral arclet, on a pattern not perfectly perpendicular to our line of
sight, therefore leading to a smaller projected velocity.

The statistical significance of the above results, and in particular
of the uncertainties on the parameters derived from the fit, depends
on the assumed uncertainty in the observed velocities, which may not
be easy to estimate. In Appendix \ref{likelihood} we discuss in detail
this dependence and how to discriminate between the different models.

\subsection{Properties of the best-fit disk model}

Figure \ref{fig:properties} shows the main physical properties of the
best-fit disk model: the surface density profile, the cumulative mass,
the equivalent thermal speed, and the aspect ratio ($h/r$), as derived
from the self-regulated disk models of \citetalias{BL99}. The
cumulative disk mass inside the outer radius of the disk is
$M_\mathrm{disk}(r_\mathrm{out})\approx (8.6\pm 0.6)
~10^6M_{\sun}$. Note that at large radii $\sigma\propto 1/r$ (also
apparent from the approximately linear growth of the cumulative mass),
with significant deviations at $r\lesssim 0.6$ pc. The equivalent
thermal speed of the self-regulated model is approximately constant at
large radii, as expected. The number density of H$_2$ molecules in the
outer disk is in the range $1-5~10^8\mbox{cm}^{-3}$, compatible with
the conditions for maser emission and consistent with models of
circumnuclear gas heated by the AGN \citep{neufeld94,pier95}. X-ray
observations \citep{bianchi01} and VLBA radio continuum observations
\citep{gallimore97} provide lower limits for the {\it electron} number
density in the disk/torus to be $n_{\mathrm e}\gtrsim 3~10^5$
cm$^{-3}$, and $n_{\mathrm e}\gtrsim 10^{6.8}$ cm$^{-3}$, respectively.

\subsection{Consistency of the self-regulated disk assumption}
\label{consistency}

Starting from Eq. (\ref{qprof}), we can compute the radius at which
$Q$ is expected to become of the order of unity in a
non-self-gravitating model. If we refer to the physical parameters
found from the fit in the previous Section (i.e.,
$M_{\bullet}=8~10^6M_{\sun}$, $\alpha=8.3~10^{-3}$, and
$\dot{M}=0.23~M_{\sun}$/yr), we find $r_Q\approx 10^{-3}$ pc. The
onset of the self-regulation mechanism is therefore expected to take
place very deep inside the disk, with respect to the radial distances
on which we focus here, so that the use of a simple completely
self-regulated model made in the previous Section is justified.

To strengthen this argument, we have also perfomed a fit by assuming
the partially self-regulated disk model, described in
\citetalias{BL99}, in which the profile of the stability parameter $Q$
is assumed to decrease according to Eq. (\ref{qprof}) for $r\ll r_Q$
and to be flat at $r\gg r_Q$, with $r_Q=10^{-3}$ pc. No significant
differences are found with respect to the completely self-regulated
case.

\section{Alternative scenarios}
\label{alt}

In this Section we discuss some possible models, alternative to the
self-gravitating disk picture, developed to explain the non-Keplerian
curve in NGC 1068. In particular, we will concentrate on the effect of
a nuclear stellar cluster and on the effect of radiation
pressure. Other models also exist (for example, the effect of a warp
in the disk), but have been discussed in the literature less
frequently.

\subsection{Nuclear stellar concentration}

One possibility is that the non-point-like source of gravitational
field be distributed spherically rather than in a disk. A spheroidal
nuclear stellar cluster, for example, could produce significant
changes to the rotation curve traced by water maser emission, if its
mass enclosed within 1 pc from the central engine exceeds $10^7
M_{\sun}$ \citep{kumar99}.

Indeed, a nuclear cusp in the luminosity profile has been observed in
NGC 1068 \citep{thatte97}. Based on stellar velocity dispersion
measurements, \citet{thatte97} estimate the dynamical mass within
$1\arcsec$ ($\approx 69.8$ pc at a distance of 14.4 Mpc) to be
$\approx 6~10^8M_{\sun}$.  \citet{schinnerer00} report a value of
$10^8 M_{\sun}$ within a 25 pc diameter from the center, based on CO
kinematics. It should be noted, however, that these mass estimates
cannot be associated with the stellar cluster only, as the observed
kinematics includes also the contributions of the central black hole
and of the disk to the gravitational field.

\citet{kumar99} has extrapolated the mass profile down to small radii
of the order of 1 pc from the central engine of NGC 1068 starting from
the mass estimate of \citet{thatte97} and assuming that the stellar
cluster can be described as a singular isothermal sphere, with stellar
density $\rho\propto r^{-2}$. In this case, the resulting mass
enclosed in 1 pc would be $\approx 8.6~10^6 M_{\sun}$, therefore able
to reproduce the desired deviations from Keplerian rotation. However,
this procedure overestimates the stellar contribution in the inner
regions, because the stellar cluster is not expected to be
characterized by the singular isothermal sphere profile down to the
innermost regions. Empirically, shallower profiles are generally found
\citep{faber97}. In addition, the adiabatic growth of a black hole at
the center of a stellar cluster leads to a density profile
$\rho\propto r^{-3/2}$ \citep{cipollina94} inside the radius of
influence of the black hole $r_{bh}=GM_{\bullet}/ \sigma_{\star}^2$,
where $\sigma_{\star}$ is the stellar velocity dispersion (curiously,
if we use the value of $\sigma_{\star}$ provided by
\citealt{thatte97}, we find that $r_{bh}$ is of the order of a few
parsec, comparable to the scales at which the maser emission is
observed in NGC 1068). Recent N-body simulations of the formation of
galactic nuclei by merger of two galaxies with initially steep density
profiles ($\rho\propto r^{-2}$) \citep{milos01} have shown that a
shallow stellar cusp is left around the nucleus, with $\rho\propto
r^{-1}$ inside a break radius of the order of $10^2$ pc.

Another difficulty with the picture of a compact stellar cluster is
based on dynamical arguments. A collisional timescale for a stellar
system characterized by a velocity dispersion $\sigma_{\star}$ and
density $\rho$ is given by:
\begin{equation}
t_{coll}=\frac{\sigma_{\star}^3}{8\pi G^2 \rho m~\mbox{ln}\Lambda},
\end{equation}
where $m$ is the stellar mass and $\mbox{ln}\Lambda$ is the Coulomb
logarithm (we will assume $\mbox{ln} \Lambda=10$). The nuclear stellar
cluster should have at least a density of $\rho=10^7
M_{\sun}/\mbox{pc}^3$ in its inner regions. If we assume
$\sigma_{\star}=150$ km/sec, the value reported by \citet{thatte97},
and a stellar mass $m=M_{\sun}$, we obtain $t_{coll}\approx 6.5~10^7$
yrs. Therefore, the nuclear stellar cluster, at the high densities
required to modify the rotation curve, will be subject to collisional
effects, that may lead to the rapid dynamical evolution of the stellar
cluster itself.

To derive firm conclusions about the effect of the stellar central
concentration on the water maser rotation curve, a more detailed
knowledge of the properties of the cluster at the smallest scales
is needed.

\subsection{Radiation pressure support}

Based on radiative transfer models of a thick torus
\citep{pierkrolik92a,pierkrolik92b}, it has been shown that radiation
pressure may reduce the importance of the gravitational field of the
central object. The relative importance of this effect changes at
different radii inside the torus and could in principle lead to a
modification of the rotation curve. The main drawback of this picture
is that radiation pressure is also very efficient at reducing the
vertical gravitational field. Therefore, we would expect the disk to
be rather thick, against the observational evidence that the water
maser emission lies in a geometrically thin structure. In fact, the
model of \citet{pierkrolik92a} was intended to give a theoretical
framework for the existence of thick tori in AGNs.  A more detailed
investigation of the effect of radiation pressure on thin
configurations would therefore be needed to assess the importance of
this process in the case of NGC 1068.

\section{Conclusions}
\label{conclusion}

In this paper we have shown how the study of water maser emission in
AGNs can be a very useful tool not only to estimate the central black
hole mass, but also to study the properties of the associated
accretion disk. In particular, we have described the non-Keplerian
rotation curve in NGC 1068 in terms of a self-gravitating accretion
disk, by fitting the data of \citet{greenhill97} with the model
described in \citet{BL99}. The quality of the fit is satisfactory and
leads to $M_{\bullet}\approx M_{disk}\simeq 8~10^6 M_{\odot}$. We have
also estimated the long-sought value of the $\alpha$ viscosity
parameter, obtaining $\alpha\approx 10^{-2}$.

Our study has additional interesting consequences. Previous estimates
of $M_{\bullet}$, using a non-self-gravitating disk model, give
$M_{\bullet} \simeq 1.5~10^7M_{\odot}$, very close to the value of the
total black hole + disk mass obtained by us. The reduced black hole
mass obtained in the self-gravitating disk scenario leads to a reduced
$L_\mathrm{Edd}$ for NGC 1068, so that the Eddington ratio turns out
to be larger than previously thought. In fact, we obtain
$L_\mathrm{bol}/L_\mathrm{Edd}\simeq 0.77$. More generally, it is
important to derive firm results on $M_{\bullet}$ also in view of the
correlations recently found between $M_{\bullet}$ and the global
properties of the host galaxy, such as the
$M_{\bullet}-\sigma_{\star}$ relation
\citep{ferrarese2000,gebhardt2000}. In this context, it is worth
noting that the masses derived from water maser emission have been
considered as the most reliable estimates; NGC 1068 is indeed one of
the galaxies in the \citet{gebhardt2000} sample.

\begin{acknowledgements}
The authors wish to thank L. Greenhill for kindly providing the water
maser data, R. Ren\`o for illuminating discussions about statistical
analysis, and J. Gallimore and M. Lombardi for interesting
suggestions. This work has been partially supported by MIUR of Italy.
\end{acknowledgements}

\appendix

\section{A likelihood ratio test for the self-gravitating disk
model}

\label{likelihood}
As noted at the end of Subsection \ref{procedure}, the results of a
statistical analysis of the water maser data depend significantly on
the uncertainty assigned to the velocity data points. In this
Appendix we examine this issue and discuss how it is connected with
the general problem of discriminating between the Keplerian disk from
the self-gravitating disk model. In fact, if the uncertainty in the
velocities were sufficiently high, we would be unable to discriminate
between the two different models.

\begin{figure}
  \centerline{
  \epsfig{figure=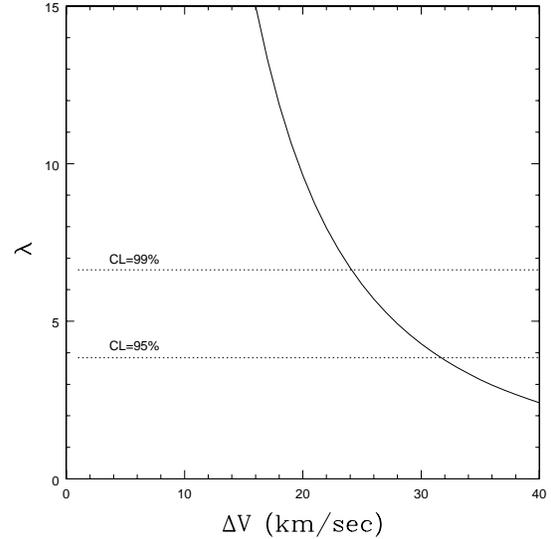,width=7.5cm}}
  \caption{Likelihood ratio $\lambda$ as a function of velocity
  uncertainty $\Delta V$. The dotted lines indicate the 95\% and the 99\%
  confidence limits.}
  \label{fig:chi2}
\end{figure}

The self-gravitating disk hypothesis is a generalization of the
Keplerian one, to which it reduces when $M_{disk}\rightarrow 0$ (or
equivalently when $r_{\mathrm s}\rightarrow\infty$). Clearly, if the
two models are fitted to the data, the minimum $\chi$-square of the
model with a larger number of parameters is going to be smaller. On
the other hand, a well-known result of statistical analysis (see
\citealt{eadie}) enables us to compare two competing hypotheses in
this case. For Gaussian, independent measurements, the ``likelihood
ratio'' $\lambda$, defined as:
\begin{equation}
\lambda=(\min\chi^2)_{\mathrm{Kep}}-(\min\chi^2)_{\mathrm{sg}},
\end{equation}
is distributed like a $\chi$-square with $n$ degrees of freedom, where
$n$ is the number of new parameters in the more general model (in our
case $n=1$), \emph{under the hypothesis that the less general model is
correct} (in our case, the Keplerian model). Of course, the resulting
value of $\lambda$ depends on the assumed uncertainties. Therefore, an
interesting question is how large should the velocity uncertainty (for
example, that associated with the magnitude of turbulent velocities)
be to make the Keplerian hypothesis acceptable with respect to the
self-gravitating one.

In Fig. \ref{fig:chi2} the likelihood ratio $\lambda$ is plotted as a
function of the assumed velocity uncertainty $\Delta V$. The two
dotted lines define the 99\% and 95\% confidence limits. The Figure
clearly shows that the Keplerian model should be rejected against the
self-gravitating one, with 99\% confidence, if $\Delta V\lesssim 25$
km/sec, and with 95\% confidence, if $\Delta V\lesssim 30$
km/sec. Note that the required uncertainty to make the Keplerian model
acceptable is more than 20 times the formal instrumental
uncertainty. This may be the reason why \citet{greenhill97} argued
that ``the scatter in the data may indicate turbulent velocities of up
to a few tens of km/sec''.

\bibliographystyle{aa}
\bibliography{lodatoaa}

\begin{thebibliography}{33}
\expandafter\ifx\csname natexlab\endcsname\relax\def\natexlab#1{#1}\fi

\bibitem[{Baan \& Hashhick(1996)}]{baan96}
Baan, W.~A. \& Hashhick, A. 1996, \apj, 269, 473

\bibitem[{Bardou {et~al.}(1998)Bardou, Heyvaerts, \& Duschl}]{bardou}
Bardou, A., Heyvaerts, J., \& Duschl, W. 1998, \aap, 337, 966

\bibitem[{Bertin \& Lin(1996)}]{bertinbook}
Bertin, G. \& Lin, C. 1996, Spiral Structure in Galaxies: a Density Wave Theory
  (Cambridge: MIT Press)

\bibitem[{Bertin \& Lodato(1999)}]{BL99}
Bertin, G. \& Lodato, G. 1999, \aap, 350, 694, (BL)

\bibitem[{Bianchi {et~al.}(2001)Bianchi, Matt, \& Iwasawa}]{bianchi01}
Bianchi, S., Matt, G., \& Iwasawa, K. 2001, \mnras, 322, 669

\bibitem[{Cipollina \& Bertin(1994)}]{cipollina94}
Cipollina, M. \& Bertin, G. 1994, \aap, 288, 43

\bibitem[{Eadie {et~al.}(1971)}]{eadie}
Eadie, W.~T. {et~al.} 1971, Statistical Methods in Experimental Physics
  (London: North Holland)

\bibitem[{Faber {et~al.}(1997)}]{faber97}
Faber, S.~M. {et~al.} 1997, \aj, 114, 1771

\bibitem[{Ferrarese \& Merritt(2000)}]{ferrarese2000}
Ferrarese, L. \& Merritt, D. 2000, \apj, 539, 9L

\bibitem[{Gallimore {et~al.}(1997)}]{gallimore97}
Gallimore, J. {et~al.} 1997, Nature, 388, 852

\bibitem[{Gallimore {et~al.}(2001)}]{gallimore01}
---. 2001, \apj, 556, 694

\bibitem[{Gebhardt {et~al.}(2000{\natexlab{a}})}]{gebhardt2000}
Gebhardt, K. {et~al.} 2000{\natexlab{a}}, \apj, 539, 13L

\bibitem[{Gebhardt {et~al.}(2000{\natexlab{b}})}]{gebhardt2000b}
---. 2000{\natexlab{b}}, \aj, 119, 1157

\bibitem[{Greenhill \& Gwinn(1997)}]{greenhill97}
Greenhill, L. \& Gwinn, C. 1997, \apss, 248, 261

\bibitem[{Greenhill {et~al.}(1996)}]{greenhill96}
Greenhill, L. {et~al.} 1996, \apj, 472, 21L

\bibitem[{Hur\'e(2000)}]{hure2000}
Hur\'e, J. 2000, \aap, 358, 378

\bibitem[{Kumar(1999)}]{kumar99}
Kumar, P. 1999, \apj, 519, 599

\bibitem[{Lin \& Pringle(1987)}]{lin87}
Lin, D. \& Pringle, J. 1987, \mnras, 225, 607

\bibitem[{Magorrian {et~al.}(1998)}]{magorrian98}
Magorrian, J. {et~al.} 1998, \aj, 115, 2285

\bibitem[{Milosavljevi\'c \& Merritt(2001)}]{milos01}
Milosavljevi\'c, M. \& Merritt, D. 2001, \apj, 563, 34

\bibitem[{Miyoshi {et~al.}(1995)}]{miyoshi95}
Miyoshi, M. {et~al.} 1995, Nature, 373, 127

\bibitem[{Neufeld {et~al.}(1994)Neufeld, Maloney, \& Conger}]{neufeld94}
Neufeld, D.~A., Maloney, P.~R., \& Conger, S. 1994, \apj, 436, L127

\bibitem[{Paczy\'nski(1978)}]{pacinski78}
Paczy\'nski, B. 1978, Acta Astronomica, 28, 91

\bibitem[{Pier \& Krolik(1992{\natexlab{a}})}]{pierkrolik92a}
Pier, E. \& Krolik, J. 1992{\natexlab{a}}, \apj, 399, 23L

\bibitem[{Pier \& Krolik(1992{\natexlab{b}})}]{pierkrolik92b}
---. 1992{\natexlab{b}}, \apj, 401, 99

\bibitem[{Pier \& Voit(1995)}]{pier95}
Pier, E. \& Voit, G.~M. 1995, \apj, 450, 628

\bibitem[{Pier {et~al.}(1994)}]{pieretal94}
Pier, E. {et~al.} 1994, \apj, 428, 124

\bibitem[{Pringle(1997)}]{pringle97}
Pringle, J. 1997, \mnras, 292, 136

\bibitem[{Schinnerer {et~al.}(2000)}]{schinnerer00}
Schinnerer, E. {et~al.} 2000, \apj, 533, 850

\bibitem[{Schmitt {et~al.}(2002)}]{schmitt02}
Schmitt, H. {et~al.} 2002, \apj, 575, 150

\bibitem[{Shakura \& Sunyaev(1973)}]{shakura73}
Shakura, N. \& Sunyaev, R. 1973, \aap, 24, 337

\bibitem[{Thatte {et~al.}(1997)}]{thatte97}
Thatte, N. {et~al.} 1997, \apj, 490, 238

\bibitem[{van~der Marel \& van~den Bosch(1998)}]{vandermarel98}
van~der Marel, R.~P. \& van~den Bosch, F. 1998, \aj, 116, 2220

\end{thebibliography}

\end{document}